\documentclass[11pt,a4paper]{article}
\usepackage{amsmath,amssymb,titling,authblk}
\usepackage[top=2.3cm,right=2.3cm,left=2.3cm,bottom=2.3cm]{geometry}
\usepackage{amssymb,amsmath,amsfonts,amsthm,amstext,amscd,array}
\usepackage{mathrsfs}
\usepackage{hyperref}
\usepackage{pdfsync}
\usepackage{bbm}
\usepackage{bm}
\usepackage[arrow,matrix,curve]{xy}
\usepackage{bbding}
\usepackage{wasysym}

\usepackage{amsfonts}
\usepackage{booktabs}
\usepackage{siunitx}

\usepackage{epsf}
\usepackage{epsfig}
\usepackage{wrapfig}
\input{epsf}

\def\ii{{\,{\rm i}\,}}
\def\dd{{\rm d}}

\newcommand{\eq}{\begin{equation}}
\newcommand{\eqend}{\end{equation}}
\newcommand{\eqa}{\begin{eqnarray}}
\newcommand{\nonueqa}{\begin{eqnarray*}}
\newcommand{\eqaend}{\end{eqnarray}}
\newcommand{\nonueqaend}{\end{eqnarray*}}

\newcommand{\bma}[1]{\begin{array}{#1}}
\newcommand{\ema}{\end{array}}
\newcommand{\bc}{\begin{center}}
\newcommand{\ec}{\end{center}}

\newcommand{\allbf}[1]{\boldsymbol{#1}}

\newif\ifold             \oldtrue

\def\bd{\begin{displaymath}}
\def\ed{\end{displaymath}}

\def\la{\label}

\newcommand{\beq}{\begin{eqnarray}}
\newcommand{\eeq}{\end{eqnarray}}

\makeatletter
\newdimen\normalarrayskip              
\newdimen\minarrayskip                 
\normalarrayskip\baselineskip
\minarrayskip\jot
\newif\ifold             \oldtrue            
\def\arraymode{\ifold\relax\else\displaystyle\fi} 
\def\@arrayskip{\ifold\baselineskip\z@\lineskip\z@
     \else
     \baselineskip\minarrayskip\lineskip2\minarrayskip\fi}
\def\@arrayclassz{\ifcase \@lastchclass \@acolampacol \or
\@ampacol \or \or \or \@addamp \or
   \@acolampacol \or \@firstampfalse \@acol \fi
\edef\@preamble{\@preamble
  \ifcase \@chnum
     \hfil$\relax\arraymode\@sharp$\hfil
     \or $\relax\arraymode\@sharp$\hfil
     \or \hfil$\relax\arraymode\@sharp$\fi}}
\def\@array[#1]#2{\setbox\@arstrutbox=\hbox{\vrule
     height\arraystretch \ht\strutbox
     depth\arraystretch \dp\strutbox
     width\z@}\@mkpream{#2}\edef\@preamble{\halign \noexpand\@halignto
\bgroup \tabskip\z@ \@arstrut \@preamble \tabskip\z@ \cr}%
\let\@startpbox\@@startpbox \let\@endpbox\@@endpbox
  \if #1t\vtop \else \if#1b\vbox \else \vcenter \fi\fi
  \bgroup \let\par\relax
  \let\@sharp##\let\protect\relax
  \@arrayskip\@preamble}
\makeatother

\newtheorem{definition}[equation]{Definition}

\theoremstyle{definition}

\usepackage{tempora}
\usepackage{color,xcolor}
\usepackage{manfnt}

\def\ddo{\end{document}}
\usepackage{hyperref}
\hypersetup{colorlinks,%
citecolor=black,%
filecolor=black,%
linkcolor=black,%
urlcolor=black}

\newcommand{\nc}{\newcommand}
\nc{\lb}{\llbracket}
\nc{\rb}{\rrbracket}
\nc{\gl}{\llbracket}
\nc{\gr}{\rrbracket}

\newcommand{\be}{\begin{equation}}
\newcommand{\ee}{\end{equation}}
\newcommand{\bea}{\begin{eqnarray}}
\newcommand{\eea}{\end{eqnarray}}

\begin{document}

\title{Charged Particle in Lie-Poisson Electrodynamics }
\author[1]{B. S. Basilio}
\author[1]{V. G. Kupriyanov}
\author[2,3]{M. A. Kurkov}
\affil[ ]{}
\affil[1]{\textit{\footnotesize CMCC-Universidade Federal do ABC, 09210-580, Santo Andr\'e, SP, 
Brazil. }}
\affil[2]{\textit{\footnotesize Dipartimento di Fisica ``E. Pancini'', Universit\`a di Napoli Federico II, Complesso Universitario di Monte S. Angelo Edificio 6, via Cintia, 80126 Napoli, Italy.}}
\affil[3]{\textit{\footnotesize INFN-Sezione di Napoli, Complesso Universitario di Monte S. Angelo Edificio 6, via Cintia, 80126 Napoli, Italy.}}
\affil[ ]{}
\affil[ ]{\footnotesize e-mail: \texttt{vladislav.kupriyanov@gmail.com, max.kurkov@gmail.com}}
\maketitle
\begin{abstract}\noindent
Lie-Poisson electrodynamics describes the semi-classical limit of non-commutative  $U(1)$  gauge theory, characterized by Lie-algebra-type non-commutativity. We focus on the mechanics of a charged point-like particle moving in a given gauge background. First, we derive explicit expressions for gauge-invariant variables representing the particle’s position. Second, we provide a detailed formulation of the classical action and the corresponding equations of motion, which recover standard relativistic dynamics in the commutative limit. We illustrate our findings by exploring the exactly solvable Kepler problem in the context of the $\lambda$-Minkowski (or the angular) non-commutativity, along with other examples.
\end{abstract}

\newpage
\section{Introduction} 
Various approaches to quantum gravity, whether based on string theory~\cite{Seiberg:1999vs} or alternative frameworks~\cite{Freidel:2005me}, converge on the conclusion that spacetime at small length scales effectively exhibits a non-commutative geometric structure. The physical implications of spacetime non-commutativity are described by non-commutative field theories~\cite{Szabo:2001kg,Szabo:2006wx}, particularly by non-commutative gauge theories~\cite{Hersent:2022gry,DimitrijevicCiric:2023hua,Meier:2023kzt,Meier:2023lku}.

Consider a non-commutative spacetime, defined as a manifold  $\mathcal{M}$ , equipped with the Kontsevich star product  $\star$ of smooth functions on it. 
The
infinitesimal gauge transformations,  $\delta^{\mathrm{nc}}_f$, on such space close the algebra:
\be
[\delta^{\mathrm{nc}}_f,\delta^{\mathrm{nc}}_g]=\delta^{\mathrm{nc}}_{-\ii [f,g]_\star}, \qquad f,g\in\mathcal{C}^{\infty}(\mathcal{M}) .
\la{fga}
\ee
 In the semi-classical limit, the $\star$-commutator 
 \be
 [f,g]_\star := f\star g - g\star f
 \ee
 is replaced by the Poisson bracket $\{f,g\}$ multiplied by the imaginary unit. In this approximation, the non-commutative algebra~\eqref{fga} simplifies to the Poisson gauge algebra \cite{Kupriyanov:2019cug,KS21}:
 \be
[\delta_{f}, \delta_{g}] = \delta_{\{f,g\}}. \la{pga}
\ee 
Poisson electrodynamics, or Poisson gauge theory, is a non-Abelian deformation of the $U(1)$ gauge theory, where the infinitesimal gauge transformations satisfy Eq.~\eqref{pga}. We refer to~\cite{Kup33} for details.

This research focuses on the Poisson gauge formalism.
 Specifically, we work with an  $n$-dimensional  flat space-time $\mathcal{M}\simeq \mathbb{R}^{n}$. The Cartesian coordinates on $\mathcal{M}$ are denoted through  $x^\mu$ ,  $\mu = 0, \dots, n-1$ . In these local coordinates, the Poisson bracket in the algebra~\eqref{pga} is defined as: 
 \be
\{f,g\} = \Theta^{\mu\nu}(x)\,\partial_{\mu}f\,\partial_{\nu} g.  
\la{pbMdef}
\ee
The Poisson bivector 
\be
\Theta^{\mu\nu}(x)  = \{x^{\mu}, x^{\nu}\}
\ee
plays the role of a deformation parameter: at the \emph{commutative limit} $\Theta\to 0$ the Poisson gauge algebra reduces to its Abelian counterpart. Therefore, throughout this article,  $\Theta^{\mu\nu}$ will be referred to as the \emph{non-commutativity parameter}.

Many non-commutativities that have attracted significant attention, such as the $\kappa$-Minkowski~\cite{Mathieu:2020ywc,Lizzi:2021rlb,Mathieu:2021mxl}, 
the $\mathfrak{su}(2)$~\cite{Hammou:2001cc,Gracia-Bondia:2001ynb,Guedes:2013vi,Vitale:2012dz}, and the $\lambda$-Minkowski (or the angular)~\cite{Lukierski:2005fc,DimitrijevicCiric:2018blz,Novikov:2019kit,DimitrijevicCiric:2019hqq,Lizzi:2021dud} ones, are of the Lie-algebra type:
\be
\{x^{\mu}, x^{\nu}\} = \mathcal{C}^{\mu\nu}_{\sigma} x^{\sigma},  \la{lpn}
\ee
where the constants $C^{\mu\nu}_\sigma$ are skew-symmetric in the upper indices and satisfy the Jacobi identity. Consequently, these parameters can be viewed as the structure constants of a Lie algebra.  The associated Poisson electrodynamics is called the Lie-Poisson electrodynamics, or the Lie-Poisson gauge theory~\cite{KKL2023}. In this article, we focus exclusively on Lie-algebra-type non-commutativities.

In the absence of charged matter, the Lie-Poisson gauge formalism for generic Lie-algebra-type non-commutativity has been fully developed at the classical level. The infinitesimal transformations of the gauge field were derived in~\cite{KS21}, the deformed  gauge-covariant field strength was constructed in~\cite{Kupriyanov:2022ohu}, and the deformed Maxwell equations were presented in~\cite{KKL2023}. General prescriptions for describing charged particles in a given gauge background were outlined in the symplectic groupoid approach~\cite{KSS}, see also \cite{DiCosmo:2023wth}. The detailed analyses of static potentials is provided in~\cite{Kupriyanov:2024dny}. The charged matter field in Poisson Electrodynamics was described in \cite{Sharapov:2024bbu}.

However, \emph{explicit} formulas for the kinematics and dynamics of a particle in arbitrary gauge backgrounds are still lacking. This paper aims to address this gap. In line with our previous works~\cite{Kup33,Kupriyanov:2022ohu,KKL2023}, the present study is based on first principles. On the kinematic side, we derive explicit expressions for the gauge-invariant variables which describe the position of the charged particle. On the dynamical side, we present explicit expressions for the gauge-invariant action and the corresponding equations of motion.

This paper is organized as follows. In Sec.~\ref{firs}, we review the relevant aspects of both standard electrodynamics and its Lie-Poisson deformation.  
In Sec.~\ref{seco}, we derive explicit expressions for gauge-invariant coordinates applicable to arbitrary Lie-algebra-type non-commutativity and provide detailed formulas for the classical action and equations of motion. Sec.~\ref{thir} focuses on  purely spatial non-commutativities. In particular, we analyze the exactly solvable $\lambda$-Minkowski Kepler problem along with other examples. The final section concludes the paper with a summary of our findings.
 
\section{From the Usual to the Lie-Poisson Electrodynamics \la{firs}}
We shall work with the Minkowski metric:
\be
\eta_{\mu\nu} = \mathrm{diag}\,(+1,-1,...,-1)_{\mu\nu} =\eta^{\mu\nu},
\ee
which will be used to raise and lower indices. 
 
\subsection*{a. Commutative case}
Consider a charged point-like particle of mass $m$ moving in a given $U(1)$ gauge background $A_{\mu}(x)$. For simplicity, we assume that the electric charge of the particle equals one.  For the purposes of this paper, it is more convenient to describe the particle’s dynamics within the Hamiltonian formalism, which preserves manifest Poincaré invariance\footnote{Of course, other Hamiltonian descriptions of the charged particle, which are not manifestly Poincaré-invariant, are possible, see e.g.\cite{LL2}.}.

 Let $\tau$ be a parameter on the world-line $x^{\mu}(\tau)$ of the particle in configuration space. From now on, an overdot denotes differentiation with respect to $\tau$, for example:
\be
\dot{x}^{\mu} = \frac{\dd x^{\mu}(\tau)}{\dd \tau}.
\ee
To simplify the notation, the explicit dependence on $\tau$ will be omitted where it does not cause confusion.

By introducing the conjugate momenta  $p_{\mu}(\tau)$ and the auxiliary variable $\Lambda(\tau)$, we can write  
the first-order action,
\be
S^{ \allbf{0}}_{\mathrm{particle}}[x,p,\Lambda] =\int \dd\tau \,\big[p_{\mu}\dot{x}^{\mu} - \Lambda\, H_{ \allbf{0}}(x,p)\big], \la{commutaction}
\ee
where
\be
H_{ \allbf{0}} := \pi_{\mu}^{\allbf{0}}\pi^{\mu}_{\allbf{0}} - m^2,
\qquad
\pi^{ \allbf{0}}_{\mu}(x,p) := p_{\mu} - A_{\mu}(x). \la{pi0def}
\ee
The subscripts and superscripts ``$ \allbf{0}$" indicate that we are dealing with the commutative case. 
Note that $\Lambda$, which acts as a Lagrange multiplier, is included as the cost of maintaining manifest Poincaré invariance in this Hamiltonian formulation.

By varying the action~\eqref{commutaction} with respect to $p_{\mu}(\tau)$, $x^{\mu}(\tau)$ and $\Lambda(\tau)$, one derives the Hamiltonian equations of motion accompanied by the constraint relation:
\be
\dot x^\mu=\Lambda\,\{x^\mu,H_{ \allbf{0}}\}_{ \allbf{0}}\,,\qquad \dot p_\mu=\Lambda\,\{p_\mu,H_{ \allbf{0}}\}_{ \allbf{0}}\,,\qquad  H_{ \allbf{0}}(x,p)=0\,, \la{eomP0}
\ee
where $\{,\}_{ \allbf{0}}$ stands for the canonical Poisson bracket on the particle's phase space,
\be
\{x^{\mu},x^{\nu}\}_{ \allbf{0}} =0,\qquad\{x^{\mu},p_{\nu}\}_{ \allbf{0}} = \delta^{\mu}_{\nu}, \qquad \{p_{\mu},p_{\nu}\}_{ \allbf{0}} =0. \la{commutatorsComm}
\ee

By adding the Maxwell action for $A_{\mu}(x)$
to the expression~\eqref{commutaction}, one arrives at the total action,  describing the dynamics of both the charged particle and the gauge field. 
This total action is invariant under the  infinitesimal gauge transformations: 
\be
\delta^{ \allbf{0}}_f A_{\mu} = \partial_{\mu} f, \quad \delta^{ \allbf{0}}_f x^{\mu}=0, \quad\delta_f^{ \allbf{0}} p_{\mu} = \partial_{\mu} f,
 \la{gaugecommut}
\ee
which close the $U(1)$ gauge algebra:
\be
[\delta_f^{ \allbf{0}}, \delta^{ \allbf{0}}_g] =0
. \la{U1alg}
\ee
The combinations $\pi_{\mu}^{\allbf{0}}$, defined in Eq.~\eqref{pi0def}, remain unchanged under these gauge transformations and will therefore be called “gauge-invariant momenta.”

Apart from that, there is another `gauge' symmetry\footnote{We placed the word `gauge' in quotation marks to emphasize the distinction between the transformations~\eqref{repar} and~\eqref{gaugecommut}.}.
Indeed, the action~\eqref{commutaction}  remains invariant under reparametrizations of the particle’s trajectory:
\be
x^{\mu}(\tau) \longrightarrow x^{\mu}(\tilde\tau(\tau)), \qquad p_{\mu}(\tau) \longrightarrow p_{\mu}(\tilde\tau(\tau)), \qquad \Lambda(\tau) \longrightarrow \Lambda_{\mu}(\tilde\tau(\tau)), \la{repar}
\ee
for any invertible and sufficiently smooth function $\tilde\tau(\tau)$.  
 By fixing a suitable `gauge', the auxiliary variable $\Lambda$ can be eliminated. We address this point in Sec.~\ref{thir}, in the context of purely spatial non-commutativities.

\subsection*{b. Lie-Poisson electrodynamics}
Our goal is to construct a suitable Lie-Poisson deformation of the Hamiltonian formalism described above.
As we mentioned in the Introduction, the starting point of the Lie-Poisson electrodynamics is the Poisson bracket~\eqref{lpn} on $\mathcal{M}$. First, we extend it to the whole phase space of the particle.

Consider an $n\times n$ momentum-dependent matrix\footnote{According to our notations, for any matrix the upper index enumerates strings, whilst the lower one enumerates columns.} $\gamma^{\mu}_{\nu}(p)$, which satisfies the \emph{first master equation}:
\be
\gamma_{ \mu }^{ \nu }(p) \,\partial^{ \mu}_p \gamma^{ \xi}_{ \lambda}(p) - \gamma^{ \xi}_{ \mu }(p)\, \partial_p^{ \mu } \gamma^{ \nu}_{ \lambda}(p) 
- \gamma^{ \mu }_{ \lambda} (p)f_{ \mu}^{ \nu  \xi} = 0, \qquad \partial_p^{\mu} := \frac{\partial}{\partial p^{\mu}}, \la{firstmaster}
\ee 
and exhibits the commutative limit:
\be
\lim_{\Theta\to 0} \gamma_{ \mu }^{ \nu } = \delta_{ \mu }^{ \nu }. \la{gammacorrlim}
\ee
Then the Poisson brackets between phase-space coordinates are defined as\footnote{A deformed symplectic structure on the phase space was also considered within a different approach in~\cite{Liang:2024tbb}.}~\cite{KS21}:
\be
\{x^{\mu},x^{\nu}\} =\Theta^{\mu\nu}(x),\qquad\{x^{\mu},p_{\nu}\}= \gamma^{\mu}_{\nu}(p), \qquad \{p_{\mu},p_{\nu}\} =0. \la{commutatorsNComm}
\ee  

 \noindent{\emph{\small {\bf Remark.} 
Let $\mathfrak{g}$ be a Lie algebra corresponding to the structure constants $\mathcal{C}^{\mu\nu}_{\sigma}$, which define the non-commutativity. In the symplectic groupoid approach to Poisson electrodynamics~\cite{KSS}, the phase space of a point-like particle is $\mathbb{R}^{n} \times G$, where G represents the $n$-dimensional Lie group associated with $\mathfrak{g}$. The correspondence between the present formalism and that of~\cite{KSS} is established by identifying our momenta $p_{\mu}$ with the local coordinates on $G$ near its identity. Then the differential operators
\be
\gamma^{\mu}(p) := \gamma^{\mu}_{\nu}(p) \, \partial_{p}^{\nu},
\ee
become the left-invariant vector fields on $G$.
}}

The infinitesimal gauge transformations, closing the deformed  algebra~\eqref{pga}, are given by:
\bea
\delta_f A_{\mu}(x) &=& \gamma^{\nu}_{\mu}\big(A(x)\big)\,\partial_{\nu} f(x) +\{A(x),f(x)\}, \nonumber \\
\delta_f x^\mu(\tau)&=&\{f(x),x^\mu\}=-{\cal C}^{\mu\nu}_\alpha x^\alpha(\tau)\,\partial_\nu f\big(x(\tau)\big)\,, \nonumber \\
\delta_f p_\mu(\tau)&=&\{f(x),p_\mu\}=\gamma_\mu^\nu\big(p(\tau)\big)\,\partial_\nu f\big(x(\tau)\big)\,.  \la{gaugenc}
\eea
At the commutative limit these equations recover the $U(1)$ transformations~\eqref{gaugecommut}.

Remarkably, for any Lie-algebra-type non-commutativity, the solution of the master equation~\eqref{firstmaster}, which satisfies the correct commutative limit~\eqref{gammacorrlim}, is known. This “universal” $\gamma$ is represented by the matrix-valued function\footnote{The subscript $\rm{\bf u}$ denotes ``universal".}~\cite{KS21}:
\be
\gamma_{\rm{\bf u}}(p) = G(\hat{p}), \qquad \hat{p}^{\mu}_{\nu} =   \mathcal{C}^{\sigma\mu}_{\nu} p_{\sigma}, \la{firsuniv}
\ee
where the form factor $G$ is defined as:
\be
G(s) = \frac{s}{2} + \frac{s}{2}\,\mathrm{coth}\,\frac{s}{2} = \sum_{k=0}^{\infty}\frac{s^k B_k^{-}}{k!},
\ee
and $B_k^{-}$, $k\in\mathbb{Z}_{+}$ are the Bernoulli numbers.

The deformed field-strength tensor $\mathcal{F}_{\mu\nu}$,
 which transforms in a gauge-covariant manner under Lie-Poisson gauge transformations,
\be
\delta_{f}\mathcal{F}_{\mu\nu}(x) = \{\mathcal{F}_{\mu\nu}(x),f(x)\}, \qquad \lim_{\Theta\to0}\mathcal{F}_{\mu\nu} = \partial_{\mu} A_{\nu} - \partial_{\nu}A_{\mu}, \la{Ftras}
\ee
can be constructed as follows~\cite{Kup33,Kupriyanov:2022ohu}:
\be
\mathcal{F}_{\mu\nu} = \rho_{\mu}^{\sigma}(A)\rho_{\nu}^{\lambda}(A)\,\big(
\gamma_{\sigma}^{\xi}(A)\partial_{\xi}A_{\lambda} -\gamma_{\lambda}^{\xi}(A)\partial_{\xi}A_{\sigma} +\{A_{\sigma},A_{\lambda}\}
\big).
\ee
In this formula, the $n \times n$ matrix $\rho^{\mu}_{\nu}(p)$ is a solution to the \emph{second master equation}\footnote{ In the symplectic groupoid approach of Ref.~\cite{KSS}, this matrix locally defines the right-invariant one-forms \(\rho_{\mu}(p) = \rho_{\mu}^{\nu}(p)\,\dd p_{\nu}\) on the group $G$; cf. the Remark.}~\cite{Kup33}, with the appropriate commutative limit:
\be
\gamma^{\nu}_{\sigma} \partial_p^{\sigma}\rho_{\alpha}^{\mu} \,+ \rho_{\alpha}^{\sigma} \partial_p^{\mu}\gamma_{\sigma}^{\nu}   = 0, 
\qquad \lim_{\Theta\to 0} \rho_{ \mu }^{ \nu } = \delta_{ \mu }^{ \nu }.
 \la{secondmaster}
\ee
The “universal” expression for $\rho$, applicable to all Lie-algebra-type non-commutativities, was constructed in~\cite{Kupriyanov:2022ohu} as follows:
\be
\rho_{\mathrm{u}}(p) = \big(\gamma_{\mathrm{u}}(p) - \hat{p}\big)^{-1} = \frac{1}{G(-\hat{p})}. \la{secouniv}
\ee

In conclusion, we note that using the transformation rules~\eqref{gaugenc} for  $A_\mu(x)$  and  $x^\mu(\tau)$, we find that the gauge field evaluated at the particle’s position
 transforms as follows:
\be
\delta_f A_{\sigma}\big(x(\tau)\big) = \gamma_{\sigma}^{\nu}\big(A\big(x(\tau)\big)\big)\,\partial_{\nu}f\,\big(x(\tau)\big). \la{picondi}
\ee 
The deformed field-strength tensor, evaluated at the position of a point-like particle, is gauge-invariant:
\be
\delta_{f}\mathcal{F}_{\mu\nu}(x(\tau)) = 0.
\ee
This can be easily verified by combining the transformation rules~\eqref{Ftras} for the field $\mathcal{F}_{\mu\nu}(x)$ with those for the particle’s coordinates $x^{\mu}(\tau)$.

In the next section, we consider the gauge-invariant kinematic and dynamical aspects of a point-like particle.

\section{Lie-Poisson gauge-invariant mechanics\la{seco}}

\subsection*{a. Gauge-invariant momenta}
As we shall soon see, the deformed gauge-invariant momenta  $\pi_\mu(p,A)$ , which generalize the commutative expressions~\eqref{pi0def}, play a key role in the construction of gauge-invariant dynamics.
These quantities are defined in such a  way that:
\be
\delta_f \pi_{\mu} = 0, \qquad \lim_{\Theta\to0}\pi_{\mu} = \pi^{\allbf{0}}_{\mu}. \la{pinv}
\ee
Using Eq.~\eqref{picondi}, we derive:
\be
\delta_f \pi_{\mu}\big(p(\tau), A(x(\tau))\big) = \big(\gamma_{\sigma}^{\nu}(A)\, \partial_{A}^{\sigma}\pi_{\mu} 
+ \gamma_{\sigma}^{\nu}(p)\, \partial_{p}^{\sigma}\pi_{\mu}
\big)\,\partial_{\nu}f. 
\ee
The condition~\eqref{pinv} implies that $\pi$ can be constructed by solving  the partial differential equation:
\be
\gamma_{\sigma}^{\nu}(A)\, \partial_{A}^{\sigma}\pi_{\mu} 
+ \gamma_{\sigma}^{\nu}(p)\, \partial_{p}^{\sigma}\pi_{\mu} = 0, \la{mepi1}
\ee
with the additional requirement:
\be
\lim_{\Theta\to 0}\pi_{\mu}(p,A) = p_{\mu} - A_{\mu}. \la{comlim}
\ee
The perturbative solution  reads,
\begin{eqnarray}
\pi_\alpha(p,A)=p_\alpha-A_\alpha+\frac{1}{2}\, {\cal C}^{\nu\beta}_\alpha\, A_\nu\, p_\beta+\frac{1}{12}{\cal C}^{\beta\mu}_\alpha{\cal C}^{\nu\sigma}_\mu\left(p_\beta p_\sigma A_\nu-A_\beta A_\sigma p_\nu\right)+{\cal O}({\cal C}^3)\,.\notag
\end{eqnarray}

Although exact solutions of these equations in the general case are not currently available, we analyze the most significant four-dimensional cases of non-commutativity. Gauge-invariant momenta for the $\kappa$-Minkowski case are presented in Sec. 3.{\bf{c}}, while the  $\mathfrak{su}(2)$  and the $\lambda$-Minkowski cases are addressed in Sec.~\ref{thir}.

\subsection*{b. Gauge-invariant dynamics}

The quantities $\pi_{\mu}$   introduced above allow the definition of a gauge-invariant deformation of the commutative Hamiltonian $H_{(\allbf{0})}$:
\be
H := \pi_{\mu}\pi^{\mu}  -  m^2.  \la{Hdefo}
\ee
First, we construct a classical action that yields the equations of motion:
\be
\dot x^\mu=\Lambda\,\{x^\mu,H\}\,,\qquad \dot p_\mu=\Lambda\,\{p_\mu,H\}\,,\qquad  H(x,p)=0\,. \label{eomP}
\ee
After that, we verify its gauge invariance.

To carry out the first part of this program, we introduce the Darboux coordinates  $X$ and $P$  on the phase space:
\be
\{X^{\mu},X^{\nu}\} =0,\qquad\{X^{\mu},P_{\nu}\} = \delta^{\mu}_{\nu}, \qquad \{P_{\mu},P_{\nu}\} =0. \la{commutatorsDarRel}
\ee
These coordinates can be constructed explicitly as:
\begin{equation}
X^\mu(x,p)=x^\nu\,\bar\gamma^\mu_\nu(p)\,, \qquad P_\mu(x,p)=p_\mu, \la{Darb}
\end{equation}
where $\bar{\gamma}$  is the inverse of the matrix $\gamma$.
It is straightforward to verify that these expressions satisfy the required Poisson bracket relations by using the equation\footnote{In the symplectic groupoid approach of Ref.~\cite{KSS}, the quantities 
$ \bar\gamma_{\mu}(p) = \bar\gamma_{\mu}^{\nu}(p) \,\dd p_{\nu} $ correspond to the left-invariant one-forms on  $G$ ; see the Remark in Sec.\ref{seco}. From this perspective, the relation~\eqref{gammabar} is the Maurer–Cartan equation.} :
\be
\partial_p^{\nu}\bar{\gamma}^{\alpha}_{\mu}(p) - \partial_p^{\alpha}\bar{\gamma}^{\nu}_{\mu}(p)
+\bar\gamma^{\nu}_{\xi}(p)\,\bar\gamma^{\alpha}_{\varepsilon}(p)\, \mathcal{C}_{\mu}^{\xi\varepsilon}  =0, \la{gammabar}
\ee
which is simply the first master equation for  $\gamma$ , rewritten in terms of the inverse matrix  $\bar{\gamma}$.

The classical action:
\be
{S}_{\mathrm{particle}}[X,P,\Lambda]=\int\! d\tau \left[P_\mu  \,\dot X^\mu -\Lambda\,H\right], 
\ee
naturally leads to the equations of motion
\be
\dot G =\Lambda\,\{G,H\}\,,\qquad  H =0, 
\ee
for any smooth function $G(X,P)$ on the phase space. By setting  $G =x^{\mu}(X,P)$ and $G= p_{\mu}(X,P)$, we recover the required equations of motion~\eqref{eomP}.

Returning to the original coordinates we obtain:
\be
S_{\mathrm{particle}}[x,p,\Lambda]=-\int d\tau\left [\dot p_\nu \bar\gamma^{\nu}_{\alpha}(p)\,x^\alpha+\Lambda\, H\right].  \label{Sginv}
\ee
This action reduces to the standard relativistic action~\eqref{commutaction} in the commutative limit. Moreover, it remains invariant under Poisson gauge transformations~\eqref{gaugenc} of the phase-space variables $x$, $p$ and the vector field $A$ .
Indeed, the gauge invariance of  $\pi_\mu$  ensures that  $\delta_f H = 0$. Consequently, the first-order Lagrangian in the action:
\be
\mathcal{L}(x,p,\Lambda)=-\dot p_\nu \bar\gamma^{\nu}_{\alpha}(p)\,x^\alpha-\Lambda\, H,
\ee
transforms as:
\bea
\delta_f \mathcal{L} &=& \frac{\dd}{\dd \tau} \left[f(x) - x^{\beta}\partial_\beta f(x) \right]  \nonumber\\
&+& \dot{p}_{\nu}\,\gamma^{\beta}_{\alpha}(p)\, x^{\mu} \,\partial_{\beta} f(x)\,\left[
\partial_p^{\nu}\bar{\gamma}^{\alpha}_{\mu}(p) - \partial_p^{\alpha}\bar{\gamma}^{\nu}_{\mu}(p)
+\bar\gamma^{\nu}_{\xi}(p)\,\bar\gamma^{\alpha}_{\varepsilon}(p)\, \mathcal{C}_{\mu}^{\xi\varepsilon}
\right]. \label{dLa}
\eea
Taking into account the equation (\ref{gammabar}), we conclude that  $\delta_f \mathcal{L}$  simplifies to a total derivative, and we obtain the desired gauge invariance:
\be
\delta_f S[x,p,\Lambda] = 0.
\ee

Summarising, the presence of non-commutativity influences the dynamics~\eqref{eomP} in two ways. First, it introduces the deformed Poisson bracket, defined by the matrix $\gamma$ through the relations~\eqref{commutatorsNComm}. The inverse matrix $\bar\gamma$ enters the classical action~\eqref{Sginv}. Second, the gauge-invariant momenta $\pi_{\mu}$, which appear in the Hamiltonian~\eqref{Hdefo}, are deformed. Below, we illustrate our formalism by presenting the main elements for the $\kappa$-Minkowski non-commutativity.

\subsection*{c. Example: $\kappa$-Minkowski non-Commutativity}
In the case of the $\kappa$-Minkowski non-commutativity, the Poisson bivector is defined as:
\be
\Theta^{\mu\nu}(x) = \kappa\,\big(\delta_0^{\mu}\, x^{\nu} - \delta_0^{\nu}\,x^{\mu}\big),
\ee
where $\kappa$ is a deformation parameter of the dimension of a length.

The matrix $\gamma$, which enters the equations of motion through the deformed Poisson bracket~\eqref{commutatorsNComm}, is given by~\cite{Kupriyanov:2022ohu}:
\begin{equation}
\gamma(p)  =  \left(
\begin{array}{cccc}
1 &-\kappa p_1 &-\kappa p_2 & -\kappa p_3 \\
0 &1 &0 &0 \\
0 &0 &1 &0 \\
0 &0 &0 &1 
\end{array}
\right).
\end{equation}
The inverse of $\gamma$  appearing in the action~\eqref{Sginv} is:
\be
\bar\gamma(p)  =  \left(
\begin{array}{cccc}
1 & \kappa p_1 & \kappa p_2 &  \kappa p_3 \\
0 &1 &0 &0 \\
0 &0 &1 &0 \\
0 &0 &0 &1 
\end{array}
\right).
\ee
The gauge-invariant momenta, which enter the Hamiltonian, can be chosen as follows:
\begin{equation}
\pi_{\mu}(p,A) =\big(\omega\, \rho_{\mu}^{\nu}(A) + (1-\omega)\, \rho_{\mu}^{\nu}(p)\big)(p_{\nu}-A_{\nu}),\quad\quad \forall \omega\in\mathbb{R}, \la{pikappa}
\end{equation}
where the matrix
\be
\rho(p)  =  \left(
\begin{array}{cccc}
1 &0 &0 & 0 \\
0 & e^{\kappa p_0} &0 &0 \\
0 &0 &e^{\kappa p_0} &0 \\
0 &0 &0 &e^{\kappa p_0} 
\end{array}
\right),
\ee
is a solution of the second master equation~\eqref{secondmaster}. 
A straightforward calculation shows that this one-parameter family of  $\pi_\mu$ satisfies the gauge invariance condition~\eqref{mepi1} and has correct commutative limit~\eqref{comlim}.

\subsection*{d. Gauge-invariant coordinates}
A peculiar feature of our framework, as highlighted by Eq.~\eqref{gaugenc}, is the gauge dependence of the particle’s coordinates.
This raises the question of whether gauge-invariant variables can be associated with the particle’s position.  

Below, we demonstrate that it is indeed possible to derive gauge-invariant deformations of the coordinates  $x^\mu$  for any  background field  $A_\mu(x)$ . 
These deformations, denoted by  $\xi^\mu(x, A(x))$, must satisfy:
\be
 \delta_f \xi^{\mu}(\tau) = 0 , \qquad \lim_{\Theta\to 0} \xi^{\mu}(\tau) = x^{\mu}(\tau).\la{invcoord}
\ee
The correct commutative limit suggests the ansatz:
\be
\xi^\mu\big(x,A(x)\big)=\Delta^\mu_\nu \big(A(x)\big)\,\,x^\nu\,, \qquad \lim_{\Theta\to 0}\Delta^\mu_\nu(p)=\delta^\mu_\nu.  \la{ginvcoordsGen}
\ee
The requirement of gauge invariance~\eqref{invcoord} 
 leads to the partial differentiall equation for  $\Delta^\mu_\nu(p)$ :
\begin{equation}
{\cal C}^{\alpha\beta}_\nu\,\Delta^\mu_\alpha (p)+\gamma^\beta_\alpha(p)\,\partial^\alpha_p\Delta^\mu_\nu(p)=0\,.
\end{equation}
It is straightforward to verify that the combination,
\begin{equation}
\Delta^\mu_\nu(p)=\bar\rho^\mu_\alpha(p)\,\bar\gamma^\alpha_\nu(p)\,, \la{DeltaSol}
\end{equation}
is a suitable solution. In this formula $\bar{\rho}$ denotes the inverse of the matrix $\rho$.

By substituting the “universal” expressions~\eqref{firsuniv} and~\eqref{secouniv} for  ${\rho}$  and  $\gamma$  into Eq.~\eqref{DeltaSol}, we express  $\Delta$  as a  matrix-valued function:
\be
\Delta(p)  = T(\hat{p}),
\ee
where the form factor  $T$  is defined as:
\be
T(s) =G^{-1}(s)\,G(-s) = \exp{(-s)}.
\ee 
Combining these results, we obtain the explicit formula for the gauge-invariant coordinates:
\be
\xi^{\mu}\big(x,A(x)\big) = \big[\exp{(- \hat{A}(x))}\big]^{\mu}_{\nu} \,x^{\nu},\qquad \hat{A}^{\mu}_{\nu}(x) := \mathcal{C}^{\sigma\mu}_{\nu} A_{\sigma}(x).
\ee

\section{Purely spatial non-commutativity: `temporal' gauge \la{thir}}
In this section, we use boldface to denote the spatial parts of  $n$-dimensional vectors and matrices, and we label the corresponding indices with Latin letters. The time variable  $x^0$  will be denoted through  $t$ . The dot~$\cdot$  denotes the standard Euclidean inner product in  
$\mathbb{R}^{n-1}$ , defined as  $\allbf{A}\cdot \allbf{B} = A^j B^j $ . For example:
\be  
 (\dot{\allbf{p}}\cdot \bar{ \allbf\gamma}\,\allbf{x}) = \dot{p}^j \bar\gamma^j_k x^k = - \dot{p}_j \bar\gamma^j_k x^k.
\ee
\subsection*{a. General case}
When the non-commutativity is purely spatial, that is,
\be
\{t, \allbf{x}\} =0,
\ee
the time components of  $\gamma(p)$,  $\rho(p)$, and $\pi(p,A)$ coincide with their commutative counterparts:
\bea
\gamma^{0}_{\mu}  &=& \delta^0_{\mu}  = \gamma_0^{\mu}, \nonumber \\
\rho^{0}_{\mu}  &=& \delta^0_{\mu}  = \rho_0^{\mu}, \nonumber \\
\pi_0  &=& p_{0} - A_0. \label{simple1}
\eea
The spatial components of these objects do not depend on $p_0$  and $A_0$, 
\be
{\allbf\gamma}  = {\allbf\gamma}(\allbf{p}),\qquad {\allbf\rho}  = {\allbf\rho}(\allbf{p}),\qquad \allbf{\pi}  = \allbf{\pi}\big(\allbf{p},\allbf{A}(t,\allbf{x})\big). \label{simple2}
\ee

Consider the \emph{temporal} `gauge'  $\tau = t$ . Performing integration by parts in the first term of the action~\eqref{Sginv} and noting that  $\dot{t} = 1$, we obtain:
\be
S_{\mathrm{particle}}[\allbf{x},p,\Lambda] = \int \dd t \, \big[p_0 + (\dot{\allbf{p}}\cdot \bar{ \allbf\gamma}\,\allbf{x})   - \Lambda\cdot \big((p_0 - A_0(t,\allbf{x}))^2 - |\allbf{\pi}|^2- m^2\big) \big]. \la{actinte}
\ee
Since  $p_0$  and  $\Lambda$  enter the integrand without their derivatives, these variables are not dynamical. The equations of motion:
\be
\frac{\delta S_{\mathrm{particle}}}{\delta p_0} = 0, \qquad \frac{\delta S_{\mathrm{particle}}}{\delta \Lambda} = 0,
\ee
imply: 
\be
p_{0}  = A_0(t,\allbf{x})\pm\sqrt{m^2 + |\allbf{\pi}|^2}. \label{constre}
\ee

Substituting this relation with the positive root  into the action~\eqref{actinte} yields the first-order action, depending only on the dynamical variables  $\allbf{x}$  and  
$\allbf{p}$:
\be
\mathcal{S}_{\mathrm{particle}}[\allbf{x}, \allbf{p}] = \int \dd t \big[(\dot{\allbf{p}}\cdot \bar{ \allbf\gamma}\,\allbf{x})+ \mathcal{H}(\allbf{x}, \allbf{p}) \big], \la{ActionSpace}
\ee
where the Hamiltonian is given by:
\be
\mathcal{H}(\allbf{x}, \allbf{p}) = \sqrt{m^2 + |\allbf{\pi}|^2} + A_0(t, \allbf{x}). \la{LLHami}
\ee
The resulting equations of motion:
\be
\dot{\allbf{x}} = \{ \mathcal{H}, \allbf{x} \}, \qquad \dot{\allbf{p}} = \{ \mathcal{H}, \allbf{p} \}, \la{Hami}
\ee
no longer involve any auxiliary variables.

In the non-relativistic approximation  $|\allbf{\pi}| \ll m$, we have:
\be
 \sqrt{m^2 + |\allbf{\pi}|^2} \simeq m + \frac{|\allbf{\pi}|^2}{2m} + \mathcal{O}\left(\frac{|\allbf{\pi}|^4}{m^4}\right), 
\ee
so the Hamiltonian simplifies to:
\be
\mathcal{H}(\allbf{x},\allbf{p}) =  \frac{|\allbf{\pi}|^2}{2m}  +  A_0(t, \allbf{x}), \label{LLHamiNR}
\ee
where the constant term  $m$  has been omitted as it is irrelevant to the dynamics.

\noindent{\emph{\small {\bf Remark.}  
The action~\eqref{ActionSpace} could have been chosen as the starting point 
for any purely spatial non-commutativity. Moreover, it is possible to consider a more general expression for the Hamiltonian:
\be
\mathcal{H}(\allbf{x},\allbf{p}) =  J(\allbf{\pi})  +  A_0(t, \allbf{x}), \la{LLHamiJ}
\ee 
where  $J$  is any sufficiently smooth function that ensures the correct commutative limit:
\be
\lim_{\Theta\to 0} J(\allbf{\pi})  = \sqrt{m^2 + |\allbf{p}-\allbf{A}|^2}. 
\ee
}}

Below we discuss two important examples of purely spatial non-commutativities: the $\mathfrak{su}(2)$ case and the $\lambda$-Minkowski case. 

\subsection*{b. $\mathfrak{su}(2)$ non-commutativity} 
The  $\mathfrak{su}(2)$  non-commutativity is characterized by the following Poisson brackets between spatial coordinates:
\begin{equation}
\{x^k,x^l\}=2\,\alpha\,{\varepsilon^{klm}}x^m,\label{su2}
\end{equation}
where  $\varepsilon^{klm}$  is the Levi-Civita symbol, and  $\alpha$  is the dimensional deformation parameter.

By calculating the matrix-valued function  $\gamma$  in~\eqref{firsuniv}, one obtains the following expression (see~\cite{Kupriyanov:2022ohu} for details):
\be
\gamma^{k}_{a}(p)=b\left(\alpha^2|\allbf{p}|^2\right)\delta^k_a-\alpha^2\chi\left(\alpha^2|\allbf{p}|^2\right)p_a \,p^k-\alpha\,\varepsilon^{akl}p_l\,, \la{gammasu2a}
\ee
where the form factors  $b(t)$  and  $\chi(t)$  are defined as:
\begin{equation}\label{chi}
\chi(t)=\frac1t\,\left(\sqrt{t}\cot\sqrt{t}-1\right)\,,\qquad\mbox{and}\qquad b(t)=1+t\,\chi(t)=\sqrt{t}\cot\sqrt{t}\,.
\end{equation}

The solutions of Eq.~\eqref{mepi1} can be chosen as follows:
\begin{eqnarray}\label{tildepi}
\allbf{\pi}&=&\frac{\arcsin{\sqrt{\alpha^2\allbf{\bar\pi}^2}}}{\sqrt{\alpha^2\allbf{\bar\pi}^2}}\,\allbf{\bar\pi}\,,\\
\allbf{\bar\pi}&=&\allbf{p}\,\frac{\sin\sqrt{\alpha^2|\allbf p|^2}}{\sqrt{\alpha^2|\allbf p|^2}}\cos\sqrt{\alpha^2|\allbf{A}|^2}-\allbf{A}\,\frac{\sin\sqrt{\alpha^2|\allbf{A}|^2}}{\sqrt{\alpha^2|\allbf{A}|^2}}\cos\sqrt{\alpha^2|\allbf{p}|^2}\notag\\
&&+\alpha\,\allbf{A}\times \allbf{p}\,\frac{\sin\sqrt{\alpha^2|\allbf{A}|^2}}{\sqrt{\alpha^2|\allbf{A}|^2}}\,
\frac{\sin\sqrt{\alpha^2|\allbf{p}|^2}}{\sqrt{\alpha^2|\allbf{p}|^2}}\,.\notag
\end{eqnarray}
The crosscheck is straightforward.
The detailed derivation of this formula is provided in Appendix~{\bf A}.

To connect the  expression~\eqref{tildepi} with earlier results on classical mechanics in non-commutative spaces~\cite{Kupriyanov:2024dny}, we perform a change of variables:
\be
\allbf{k}=\allbf{ p}\,\frac{\sin\sqrt{\alpha^2|\allbf{p}|^2}}{\sqrt{\alpha^2|\allbf{p}|^2}}\,,
\ee
which yields the following solution of the first master equation~\eqref{firstmaster}:
\be
\tilde\gamma^i_j(k)=\sqrt{1-\alpha^2|{ \allbf{k}}|^2} \,\delta^i_{j}-\alpha\,\varepsilon^{jik}\,k_k\,.
\ee
The corresponding gauge-invariant momenta are given by:
\be
\allbf{\tilde\pi}(\allbf{k},\allbf{A})=\sqrt{1-\alpha^2 |\allbf{A}|^2} \,\,\allbf{k}- \sqrt{1-\alpha^2 | \allbf{k}|^2}\,\,\allbf{A}+\alpha\,\allbf{k}\times\allbf{A}, \la{pisu2}
\ee
where the cross product is defined in the usual three-dimensional sense. This expression satisfies the gauge-invariance condition~\eqref{mepi1} and  exhibits the correct commutative limit at $\alpha \to 0$.

By setting
\be
J(\allbf{\tilde\pi}) = \frac{1}{\alpha^2 m} - \frac{\sqrt{1-\alpha^2 |\allbf{\tilde\pi}|^2}}{\alpha^2 \,m},
\ee
in Eq.~\eqref{LLHamiJ}, we obtain the non-relativistic Hamiltonian:
\be
\mathcal{H}(\allbf{x},\allbf{k}) =  \frac{1}{\alpha^2 m} - \frac{\sqrt{1-\alpha^2 |\allbf{\tilde\pi}|^2}}{\alpha^2 \,m} +  A_0(t, \allbf{x}). \la{su2NewHam}
\ee
This Hamiltonian is of particular interest because it leads to the super-integrable Kepler problem~\cite{Kupriyanov:2024dny} when the Coulomb gauge potential:
\be
\allbf{A} = \allbf{0},\qquad A_0 = \frac{Q}{|\allbf{x}|}, \qquad Q\in\mathbb{R}.\la{Coul}
\ee
is considered.
For detailed analysis of the integrals of motion and the numerical solutions of the equations of motion, refer to the cited reference.

\subsubsection*{c. $\lambda$-Minkowski non-commutativity} 
For the $\lambda$-Minkowski (also known as the  ``angular") non-commutativity, the only non-zero Poisson brackets between the spatial coordinates are:
\begin{equation}
\{{x}^3,{x}^1\} = - \lambda {x}^2 =-\{{x}^1,{x}^3\}, \quad \{{x}^3,{x}^2\} = \lambda {x}^1 =-\{{x}^2,{x}^3\}, 
\end{equation}
where $\lambda$ is a dimensional deformation parameter. 

The spatial parts  $\allbf{\gamma}$ and $\allbf{\rho}$ of the solutions of the master equations~\eqref{firstmaster} and~\eqref{secondmaster}
can be chosen as follows  (see~\cite{Kupriyanov:2022ohu} for details):
\begin{equation}
{\allbf{\gamma}}(\allbf{p}) = \left( \begin {array}{ccc} 
\noalign{\medskip}  
1&0&0\\ 
\noalign{\medskip}
 0&1&0
\\ 
\noalign{\medskip} 
-\lambda\,p_{{2}}&\lambda\,p_{{1}}&1\end {array}
 \right), 
 \qquad {\allbf{\rho}}(\allbf{p}) = \left( \begin {array}{ccc} 
 \noalign{\medskip}  
 \cos \left( p_{{3}}\lambda \right) &-\sin
 \left( p_{{3}}\lambda \right) &0
 \\ \noalign{\medskip} 
 \sin \left( p_{{3}}\lambda \right) &\cos \left( p_{{3}}\lambda \right) &0
\\ \noalign{\medskip} 
0&0&1\end {array} \right).  \la{gammarholambda}
\end{equation}
A one-parameter family of gauge-invariant momenta that satisfiy Eq.~\eqref{mepi1} is given by:
\begin{equation}
\pi_{j}(p,A) =\big(\omega\, \rho_{j}^{k}(A) + (1-\omega)\, \rho_{j}^{k}(p)\big)(p_{k}-A_{k}),\quad\quad \forall \omega\in\mathbb{R}. \la{pilambda}
\end{equation}
To simplify the forthcoming analysis, we set $\omega = 1$.

Consider the non-commutative dynamics of a particle in a Coulomb gauge potential~\eqref{Coul}.
At $|\allbf{x}|\neq 0$, this gauge configuration satisfies the deformed Maxwell equations, derived in Ref.~\cite{KKL2023}.
The corresponding gauge-invariant momenta $\allbf{\pi}$ coincide with the usual momenta $\allbf{p}$, and the non-relativistic Hamiltonian~\eqref{LLHamiNR} reduces to its commutative counterpart:
\be
\mathcal{H}(\allbf{x},\allbf{p}) = \frac{|\allbf{p}|^2}{2m}+\frac{Q}{|\allbf{x}|}. \label{HamCoul}
\ee

Let us introduce the new variables $X^j$ and $P_k$ as follows:
\be
\allbf{X} =\bar{\allbf{\rho}}(\allbf{p})\,\allbf{x},\qquad \allbf{P} =\bar{\allbf{\rho}}(\allbf{p})\,\allbf{p},
\ee 
with
\be
\bar{\allbf{\rho}}(\allbf{p}) = \left( \begin {array}{ccc} 
 \noalign{\medskip}  
 \cos \left( p_{{3}}\lambda \right) &\sin
 \left( p_{{3}}\lambda \right) &0
 \\ \noalign{\medskip} 
- \sin \left( p_{{3}}\lambda \right) &\cos \left( p_{{3}}\lambda \right) &0
\\ \noalign{\medskip} 
0&0&1\end {array} \right). 
\ee
A straightforward calculation shows that $X^j$ and $P_k$ are the Darboux coordinates on the phase space:
\be
\{X^{j},X^{k}\} =0,\qquad\{X^{j},P_{k}\} = \delta^{j}_{k}, \qquad \{P_{j},P_{k}\} =0. \la{commutatorsDar}
\ee
Remarkably, the Hamiltonian~\eqref{HamCoul} remains unchanged under the transformation $(\allbf{x},\allbf{p})\to(\allbf{X},\allbf{P}) $. Therefore, the equations of motion in terms of the new variables:
\be
\dot{\allbf{X}} = \{ \mathcal{H}, \allbf{X} \}, \qquad \dot{\allbf{P}} = \{ \mathcal{H}, \allbf{P} \}, \qquad
\mathcal{H}(\allbf{X},\allbf{P}) = \frac{|\allbf{P}|^2}{2m}+\frac{Q}{|\allbf{X}|},\la{HamiKeplNewC}
\ee
are identical to the commutative case. 

\begin{figure}[t]
\begin{minipage}[h]{0.49\linewidth}
\center{\includegraphics[width=1\linewidth]{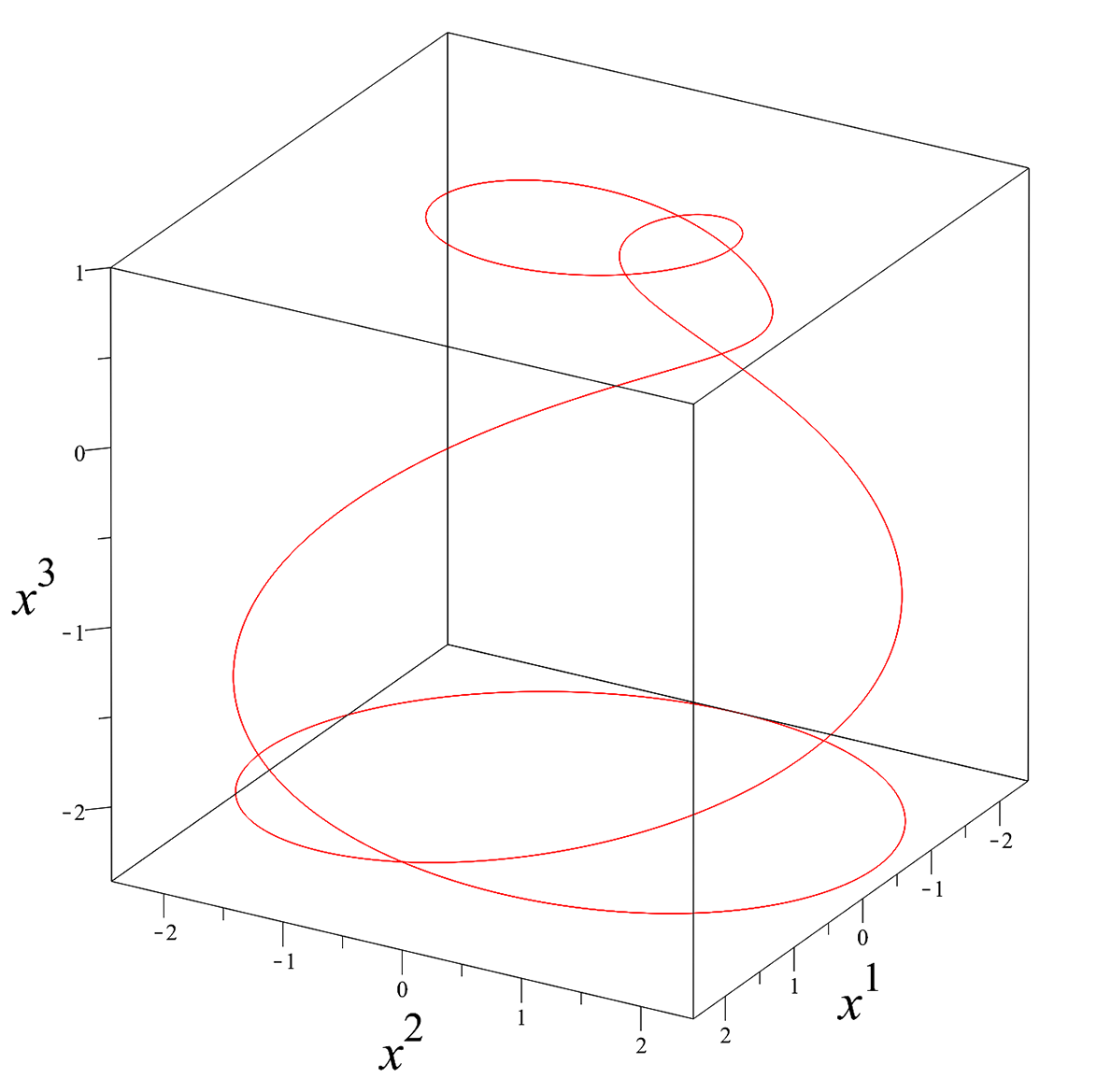}}  \\
\end{minipage}
\hfill
\begin{minipage}[h]{0.49\linewidth}
\center{\includegraphics[width=1\linewidth]{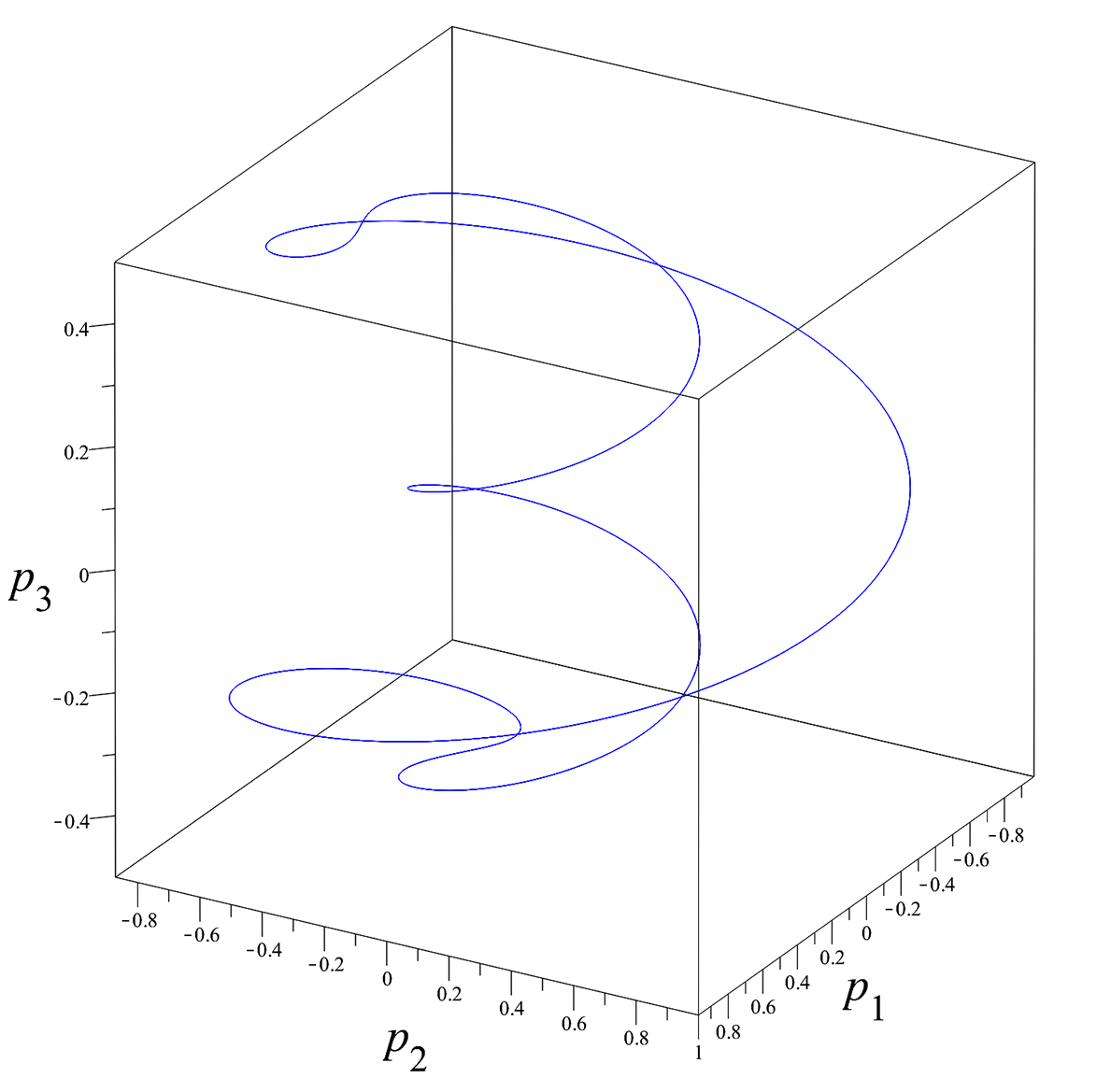}}  \\
\end{minipage}
\caption{\sl 
The $\lambda$-Minkowski case: numerical solutions of the Hamilton equations~\eqref{Hami} for the non-relativistic Hamiltonian~\eqref{LLHamiNR} with the Coulomb gauge potential~\eqref{Coul} are presented by the red and the blue lines.   
The parameters and the initial conditions are chosen as follows: $m = 1$, $\lambda = 10$, $ Q = -1$,
$x^1 (0) = 1$, $x^2 (0)=0$, $x^3 (0)=1$, $p_1(0)=0$, $p_2 (0)=1$ and $p_3(0)=1$.
 \label{fig2}
}
\end{figure}

Let $ \allbf{X}(t) $ and $\allbf{P}(t)$ be solutions of the commutative Kepler problem~\eqref{HamiKeplNewC}. The corresponding non-commutative solutions of~\eqref{Hami} can then be obtained through the inverse transformation:
\be
\allbf{x}(t) ={\allbf{\rho}}\big(\allbf{P}(t)\big)\,\allbf{X}(t),\qquad \allbf{p}(t) ={\allbf{\rho}}\big(\allbf{P}(t)\big)\,\allbf{P}(t), \la{NCsols}
\ee
where, we remind, the matrix $\allbf{\rho}$ is given by Eq.~\eqref{gammarholambda}. 
Typical trajectories are shown in Fig.~\ref{fig2}.

Consider the deformed angular momentum $\allbf{L}$ and the deformed Laplace–Runge–Lenz vector $\allbf{R}$, which Poisson-commute with the Hamiltonian~\eqref{HamCoul}:
\be
\{\allbf{L},\mathcal{H}\}=0,\qquad \{\allbf{R},\mathcal{H}\}=0.
\ee  
These vectors equal to the rotated versions:
\be
\allbf{L} = \bar{\allbf{\rho}}(\allbf{p})\,\allbf{L}_{(\allbf{0})},\qquad \allbf{R} = \bar{\allbf{\rho}}(\allbf{p})\,\allbf{R}_{(\allbf{0})}, \la{LRdeform}
\ee
of the corresponding undeformed objects:
\be
\allbf{L}_{(\allbf{0})} =  \allbf{x} \times \allbf{p},
\qquad
 \allbf{R}_{(\allbf{0})} = \allbf{p}\times\allbf{L}_{(\allbf{0})} + \frac{Q\,\allbf{x}}{|\allbf{x}|}.
 \ee
One can easily see that:
\be
\big|\allbf{L}\big|= \big|\allbf{L}_{(\allbf{0})}\big|,  \qquad L^3= L_{(\allbf{0})}^3,\qquad \big|\allbf{R}\big|= \big|\allbf{R}_{(\allbf{0})}\big|,  \qquad R^3= R_{(\allbf{0})}^3. 
\ee

The relations:
\be
\allbf{L}\cdot \allbf{R} = 0,\qquad |\allbf{R}|^2 = Q^2m^2 + 2\,m\,|\allbf{L}|^2\,\mathcal{H},
\ee
between these integrals of motion, and their algebra with respect to the Poisson brackets,
\be
\{L^{i},L^{j}\} = \varepsilon^{ijk} \,L^k,\qquad \{R^{i},L^{j}\} = \varepsilon^{ijk} \,R^k\qquad \{R^{i},R^j\}=-2\,m\,\varepsilon^{ijk}\,\mathcal{H}\,L^k,
\ee
have exactly the same form as in the commutative case.

We notice that the relation~\eqref{NCsols}, which facilitates deriving non-commutative solutions from their commutative counterparts, is not limited to the Kepler problem. This formula is, in fact, applicable to any gauge background exhibiting axial symmetry:
\be
\allbf{A}(\allbf{O}\allbf{x},t)= \allbf{O}\allbf{A}(\allbf{x},t),\qquad A_0(\allbf{O}\allbf{x},t) = A_0(\allbf{x},t),  \la{axisymm}
\ee
where    $\allbf{O}$  denotes a general rotation matrix about the  $x^3$-axis.
Indeed, this condition ensures the equality $ \mathcal{H}(\allbf{x},\allbf{p}) = \mathcal{H}(\allbf{X},\allbf{P}) $, thereby proving the above statement.

\begin{figure}[t]
\begin{minipage}[h]{0.49\linewidth}
\center{\includegraphics[width=1\linewidth]{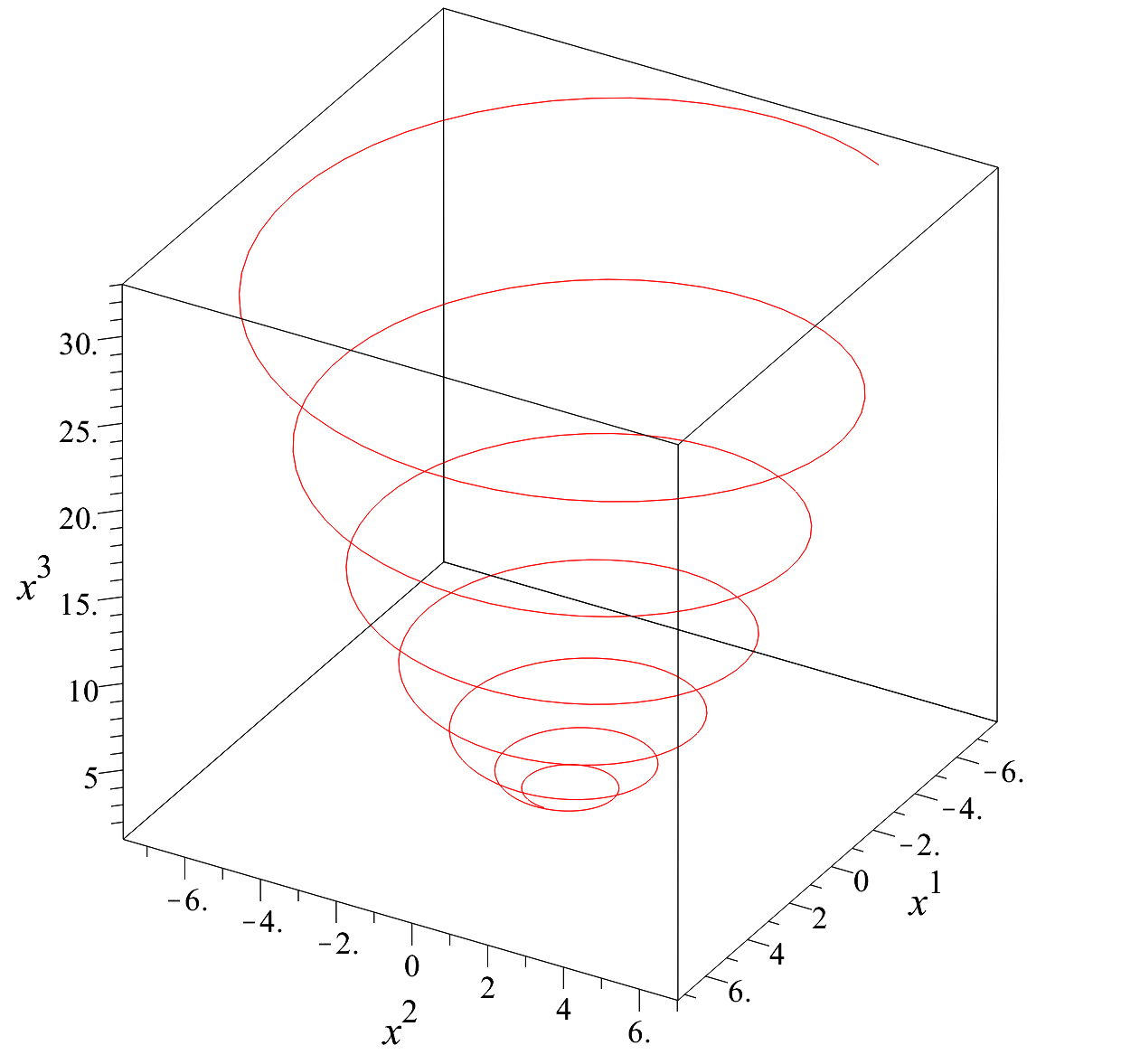}}  \\
\end{minipage}
\hfill
\begin{minipage}[h]{0.49\linewidth}
\center{\includegraphics[width=1\linewidth]{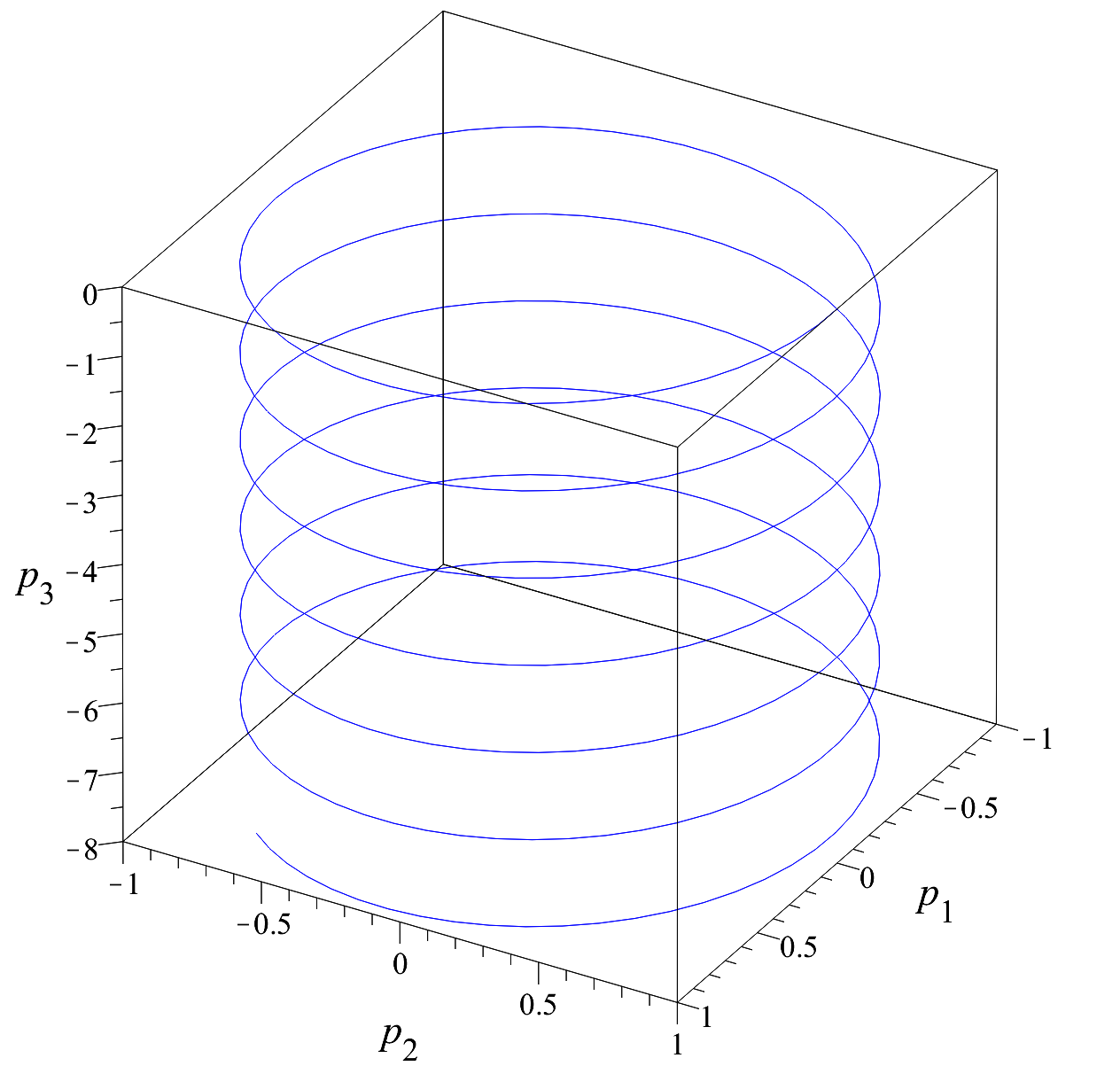}}  \\
\end{minipage}
\caption{\sl 
The $\lambda$-Minkowski case: numerical solutions of the Hamilton equations~\eqref{Hami} for the non-relativistic Hamiltonian~\eqref{LLHamiNR} are presented by the red and the blue lines.  The  gauge potential is given by~\eqref{lingb}.
The parameters and the initial conditions are chosen as follows: $m = 1$, $\lambda = 5$, $ E = -1$,
$x^1 (0) = 1$, $x^2 (0)=0$, $x^3 (0)=1$, $p_1(0)=0$, $p_2 (0)=1$ and $p_3(0)=1$.
 \label{fig3}
}
\end{figure}

For example, for the field configuration:
\be
\allbf{A} = 0,\qquad A_0 = E\,x^{3},\qquad E\in\mathbb{R}, \la{lingb}
\ee
using~\eqref{NCsols}, we can immediately derive the solutions of the Hamiltonian equations~\eqref{Hami}:
\bea
x^1(t) &=& \bigg( -  x^2(0) + \frac{p_2(0) \,t}{m} \bigg)\sin{(\lambda\,E\,t)} +\bigg(  x^1(0) - \frac{p_1(0) \,t}{m} \bigg)\cos{(\lambda\,E\,t)}, \nonumber\\
x^2(t) &=& \bigg(  + x^1(0) - \frac{p_1(0) \,t}{m} \bigg)\sin{(\lambda\,E\,t)} +\bigg(  x^2(0) - \frac{p_2(0) \,t}{m} \bigg)\cos{(\lambda\,E\,t)}, \nonumber\\
x^3(t) &=& x^3(0) - \frac{p_3(0)\,t}{m} - \frac{E\,t^2}{2m}, \nonumber\\
p_1(t) &=& \cos{(\lambda\,E\,t)}\, p_1(0) - \sin{(\lambda\,E\,t)}\,p_2(0),\nonumber\\ 
p_2(t) &=& \cos{(\lambda\,E\,t)}\, p_2(0) + \sin{(\lambda\,E\,t)}\,p_1(0),\nonumber\\ 
p_3(t) &=& p_3(0) + E\,t.
\eea
The corresponding trajectories are presented in Fig~\ref{fig3}.

However, in the general case, the formula~\eqref{NCsols} is not applicable. For instance, even a simple field configuration like:
\be
\allbf{A} = 0,\qquad A_0 = E\,x^{1},\qquad E\in\mathbb{R},
\ee
is not compatible with Eq.~\eqref{axisymm}. Nevertheless, the corresponding solutions are straightforward to derive, given the simplicity of this particular problem:
\bea
x^1(t) &=& -\frac{E\,t^2}{2\,m} - \frac{p_1(0)\,t}{m} +x^1(0), \nonumber\\
x^2(t) &=&  - \frac{p_2(0)\,t}{m} +x^2(0), \nonumber\\
x^3(t) &=&-\frac{E\,\lambda\,p_2(0)\,t^2}{2\,m}+ \frac{E\,\lambda\,x^2(0)-p_3(0)\,t}{m} +x^3(0), \nonumber\\
p_1(t) &=& E\,t + p_1(0),\nonumber\\
p_2(t) &=&  p_2(0), \nonumber\\
p_3(t) &=& p_3(0).
\eea

In conclusion, we comment on the gauge-invariant coordinates  $\xi$, introduced in Sec.~\ref{seco}. In all the examples presented above,  $\allbf{A} = \allbf{0}$, and therefore, the gauge-invariant coordinates  $\xi$  coincide with the usual coordinates $x$.

\section{Summary}

We studied the dynamics of test particles within the framework of Lie-Poisson gauge formalism. Below, we summarize the new results obtained in this paper:
\begin{enumerate}
\item \textbf{Gauge-Invariant Coordinates}: We derived the universal expression~\eqref{ginvcoordsGen} for gauge-invariant coordinates, which is valid for all Lie-algebra-type non-commutativities.
 \item \textbf{Classical Action and Equations of Motion}: We established the explicit expression~\eqref{Sginv} for the classical action and the corresponding equations of motion~\eqref{eomP}, which respect the Lie-Poisson gauge symmetry and recover standard relativistic dynamics in the commutative limit.
\item \textbf{Spatial Non-Commutativities}: For purely spatial non-commutativities, such as the $\mathfrak{su}(2)$ and the $\lambda$-Minkowski ones, we presented the classical action~\eqref{ActionSpace} and the Hamiltonian equations of motion~\eqref{Hami}, which do not involve auxiliary variables.
\item \textbf{Gauge-Invariant Momenta}: The gauge-invariant momenta $\pi_\mu$ play a key in our construciton. We provided explicit expressions for these objects for prominent Lie-algebra-type non-commutativities, namely for the $\kappa$-Minkowski~\eqref{pikappa}, for the $\mathfrak{su}(2)$~\eqref{pisu2}, and for  the $\lambda$-Minkowski~\eqref{pilambda} cases.
\item \textbf{$\mathfrak{su}(2)$ Non-Commutativity}: We presented the Hamiltonian~\eqref{su2NewHam}, which yields the super-integrable Kepler problem, discussed in Ref.~\cite{Kupriyanov:2024dny}, when the Coulomb gauge potential is considered. The expression~\eqref{su2NewHam} is valid for a generic gauge background.
\item \textbf{$\lambda$-Minkowski Non-Commutativity}: In the context of the $\lambda$-Minkowski non-commutativity, we analyzed the Kepler problem. We expressed non-commutative solutions in terms of their commutative counterparts~\eqref{NCsols}. Additionally, we presented and analyzed the deformed angular momentum and Laplace--Runge--Lenz vector~\eqref{LRdeform}.
\end{enumerate}


\subsection*{Acknowledgments}

 B.S.B. acknowledges partial support from CAPES, Brazil. V.G.K.thanks support from the National Council for Scientific and Technological Development (CNPq) Grant 304130/2021-4 and FAPESP Grant 2024/04134-6.

\section*{A. Derivation of Eq.~\eqref{tildepi}}
For the forthcoming discussion, it is convenient to introduce the notations:
\be
r:=\alpha^2|\allbf{p}|^2,\qquad s:=\alpha^2 \allbf{A}\cdot \allbf{p},  \qquad q:=\alpha^2 |\allbf{A}|^2.
\ee
The order by order  {perturbative analysis of Eq.~\eqref{mepi1}} suggests the ansatz,
\begin{equation}\label{pi1}
\allbf{\bar\pi}(\allbf{p},\allbf{A})=c\left(r,s,q\right)\left[b\left(s\right)\allbf{p}-b\left(r\right)\allbf{A}+\alpha\,\allbf{A}\times\allbf{p}\right],
\end{equation}
with the symmetry property:
\begin{equation}
c\left(r,s,q\right)=c\left(q,s,r\right), \label{symcond}
\end{equation}
and the  {commutative limit}: 
\be
\lim_{\alpha\to 0}c\left(r,s,q\right)=1. \label{ccomlim}
\ee

By substituting this ansatz and the expression~\eqref{gammasu2a} for $\gamma$ into the master equation~\eqref{mepi1} we arrive at the following four equations for $c(r, s, q)$:
\begin{eqnarray}
&&2\,\partial_q\left(c\,b(q)\right)+b(q)\left(b(r)-s\,\chi(q)\right)\partial_sc+c=0\,,\label{z1}\\
&&2\,\partial_q c+\left(b(r)-s\,\chi(q)\right)\partial_sc-\chi(q)\,c=0\,,\label{z2}\\
&&2\,\partial_r\left(c\,b(r)\right)+b(r)\left(b(q)-s\,\chi(r)\right)\partial_s c+c=0\,,\label{z3}\\
&&2\,\partial_rc+\left(b(q)-s\,\chi(r)\right)\partial_sc-\chi(r)\,c=0\,.\label{z4}
\end{eqnarray}
The definitions~\eqref{chi} of the form factors $b(q)$ and $\chi(q)$ yield the relation:
\begin{equation}
2\,\partial_q b(q)+\chi(q)b(q)+1=0\,, \label{consirel}
\end{equation}
which implies that Eq.~\eqref{z1} is equivalent to Eq.~\eqref{z2}, and Eq.~\eqref{z3} is equivalent to Eq.~\eqref{z4}.

  The simplest solution of the independent equations~\eqref{z2} and~\eqref{z4} can be obtained by assuming $\partial_s c=0$. 
 Separating the remaining two variables and imposing the symmetry condition~\eqref{symcond} along with the commutative limit~\eqref{ccomlim}, we find:
\begin{equation}
c(r,q)=\frac{\sin\sqrt{r}}{\sqrt{r}}\,\frac{\sin\sqrt{q}}{\sqrt{q}}\,.
\end{equation}
The corresponding gauge invariant momenta are given by: 
\begin{eqnarray}\label{tildepi1}
    \allbf{\bar\pi}&=&\allbf{p}\,\frac{\sin\sqrt{\alpha^2|\allbf p|^2}}{\sqrt{\alpha^2|\allbf p|^2}}\cos\sqrt{\alpha^2|\allbf{A}|^2}-\allbf{A}\,\frac{\sin\sqrt{\alpha^2|\allbf{A}|^2}}{\sqrt{\alpha^2|\allbf{A}|^2}}\cos\sqrt{\alpha^2|\allbf{p}|^2}\\
&&+\alpha\,\allbf{A}\times\allbf{p}\,\frac{\sin\sqrt{\alpha^2|\allbf{A}|^2}}{\sqrt{\alpha^2|\allbf{A}|^2}}\,
\frac{\sin\sqrt{\alpha^2|\allbf{p}|^2}}{\sqrt{\alpha^2|\allbf{p}|^2}}\,.\notag
\end{eqnarray}

By redefining the  momenta as:
\begin{equation}
\allbf{\pi}(p,A)=\frac{\arcsin{\sqrt{\alpha^2\allbf{\bar\pi}^2}}}{\sqrt{\alpha^2\allbf{\bar\pi}^2}}\,\allbf{\bar\pi}\,, \label{pi}
\end{equation}
we obtain other solutions of Eq.~\eqref{mepi1} with the correct commutative limit. These new gauge-invariant momenta exhibit a peculiar property:
\be
\allbf{\pi}|_{A=0}=\allbf{p},
\ee
which is absent for $\allbf{\bar\pi}$.

\end{document}